\documentclass[floats,floatfix,amssymb,prd,
superscriptaddress,nofootinbib,preprintnumbers]{revtex4-1}

\usepackage{natbib}
\usepackage{subcaption}
\usepackage{ragged2e}
\DeclareCaptionJustification{justified}{\justifying}
\makeatletter
\newcommand{\subsetsim}{\mathrel{\mathpalette\subset@sim\relax}}
\newcommand{\subset@sim}[2]{%
  \vtop{\offinterlineskip\m@th
    \ialign{\hfil##\cr
      $#1\subset$\cr\noalign{\kern0.5pt}\scalebox{0.9}{$#1\sim$}\cr
    }%
  }%
}
\makeatother
\usepackage{amssymb,amsmath,verbatim,mathtools,needspace,enumitem,etoolbox,graphicx,physics,microtype,afterpage,bm}
\usepackage[dvipsnames, usenames]{xcolor}
\definecolor{linkcolor}{rgb}{0.0,0.3,0.5}
\usepackage{booktabs}
\usepackage[unicode, colorlinks=true, linkcolor=linkcolor, citecolor=linkcolor, filecolor=linkcolor,urlcolor=linkcolor, pdfusetitle]{hyperref}
\usepackage[all]{hypcap}
\usepackage[T1]{fontenc}
\usepackage[utf8]{inputenc}
\usepackage{tabularx}
\usepackage{float}
\usepackage{xspace} 
\usepackage{multirow}
\usepackage{pifont}
\usepackage{lmodern}
\allowdisplaybreaks
\usepackage{tikz}
\usepackage{color}
\usepackage{framed}
\usepackage{hyperref}
\hypersetup{colorlinks, citecolor=bluscuro, linkcolor=black, urlcolor=bluscuro}
\definecolor{rossos}{cmyk}{0,1,1,0.55}
\definecolor{bluscuro}{rgb}{0.15, 0.2, .85}
\definecolor{bluchiaro}{cmyk}{1,.3,0.,0.1}
\definecolor{ForestGreen}{rgb}{0.13, 0.55, 0.13}

\def\d{{\mathrm{d}}}

\newcommand{\es}{\end{subequations}}
 
\newcommand{\be}{\begin{equation}}
\newcommand{\ee}{\end{equation}}
\renewcommand{\d}{{\rm d}}

\newcommand{\llp}{\left [}
\newcommand{\rrp}{\right ]}
\newcommand{\lp}{\left (}
\newcommand{\rp}{\right )}

\newcommand{\GI}{\text{\tiny GI}}

\newcommand{\bea}{\begin{eqnarray}}
\newcommand{\eea}{\end{eqnarray}}
\def\beq{\begin{equation}}
\def\eeq{\end{equation}}

\def\d{{\rm d}}

\def\beqa{\begin{eqnarray}}

	\def\eeqa{\end{eqnarray}}
\def\lsim{\mathrel{\rlap{\lower4pt\hbox{\hskip0.5pt$\sim$}}
		\raisE_1pt\hbox{$<$}}}         
\def\gsim{\mathrel{\rlap{\lower4pt\hbox{\hskip0.5pt$\sim$}}
		\raisE_1pt\hbox{$>$}}}         

\def\d{{\rm d}}

\def\d{{\rm d}}

\def\eeqa{\end{eqnarray}}
\numberwithin{equation}{section}

\def\bq{\begin{quote}}
\def\eq{\end{quote}}

 at 10truept

\usepackage[normalem]{ulem}

\def\wt{\widetilde}

\def\C{{\cal{C}}}

\def\cs2{c_{\rm{s}}^2}

\def\U0{{\bar U_0}}
\def\wt{\widetilde}

\newcommand\gami[1]{{\gamma_{{#1}}^{~i}}}

\newcommand\A[1]{{\phi_{{#1}}}}

\def\X{{\cal{X}}}

\def\XB{{{\cal{X}}_{\rm{B}}}}
\def\U{{\cal{U}}}

\def\syn{\text{\tiny TT}}

\newcommand{\athens}{{Physics Division, National Technical University of Athens, Athens, 15780, Greece}}

\newcommand{\unige}{D\'epartement de Physique Th\'eorique and Centre for Astroparticle Physics (CAP), Universit\'e de Gen\`eve, 24 quai E. Ansermet, CH-1211 Geneva, Switzerland}

\begin{document}
\vspace*{2cm}
\title{
On Nonlinear Black Hole Ringdowns\\
 from Gauge-Invariance and Measurements
}


\author{A. Kehagias}
\email{kehagias@central.ntua.gr}
\affiliation{\athens}
\affiliation{CERN, Theoretical Physics Department, Geneva, Switzerland}

\author{A. Riotto}
\email{Antonio.Riotto@unige.ch}
\affiliation{\unige}
\affiliation{Gravitational Wave Science Center (GWSC), Universit\'e de Gen\`eve, CH-1211 Geneva, Switzerland}

\date{\today}
\vskip 2cm

\begin{abstract}
\noindent
\begin{center}
{\bf Abstract}
\end{center}
\noindent
It has been recently shown that nonlinear effects emerging at the time of the generation of the quasi-normal modes are necessary to model  ringdowns from black hole mergers. In this note we 
describe how   nonlinerarities  also arise  when defining  gauge-invariant  tensor modes and in the calculation of the observable measured in the interferometers beyond linear order. 

\end{abstract}

\maketitle

\section{Introduction and Conclusions}\label{intro}
\noindent
First-order BH perturbation theory is standardly adopted to study the  Quasi Normal Modes (QNMs)   generated  by a perturbed Black Hole (BH) during the ringing down phase \cite{Kokkotas:1999bd,Berti:2009kk}. 
QNMs  are  determined solely by the mass, the spin and the charge of the BH  and, as such, they are fundamental in  gravitational wave astronomy. 
It has been recently showed by BH-merger simulations \cite{nl2,nl3} (see also Refs. \cite{nlold1,nlold2,Lagos:2022otp}), that not only first-order  but also second-order
effects are relevant to describe  ringdowns. 
In particular, the
nonlinear mode  amplitude arising from the square of the fundamental $(\ell,m)=(2,2)$  mode
is comparable to or it can be even larger than that of the linear mode $(4,4)$. 
Therefore,  to  correctly model the BH ringdown, the inclusion of  nonlinear effects is unavoidable. It is worth noting that, in the  case of  Kerr BHs, such nonlinearities find their explanation   in symmetry arguments \cite{Kehagias:2023ctr}.

 In this simple note we would like to describe how second-order effects arise by two other sources when dealing with gravitational wave strains. The first source arises when working with gauge-invariant second-order tensor perturbations, see also Refs. \cite{g1,g2,g3,Nakano:2007cj,g4}. Tensor modes are gauge-invariant at first-order in perturbation theory, but they are not at  second-order. 
 The construction of a gauge-invariant second-order tensor mode unavoidably introduces (first-order)$^2$ terms which are potentially relevant when dealing with 
 comparison between observation and theoretical predictions. 
 
 Unfortunately, 
 there exist infinite ways  to render  tensor modes gauge invariant at second-order, depending on the gauge one starts from. Which gauge one should adopt is in fact suggested by the measurement procedure and,  in order to give a description of the response of the detector,   the best choice seems to be the so-called TT frame \cite{book}. As we will see, the expression of the time shifts measured in interferometers in the TT gauge contains as well second-order (first-order)$^2$ terms. Furthermore, analytical calculation of the QNMs are best performed in the so-called Regge-Wheeler (RW)  gauge where   
Schwarzschild perturbations are solved through the RW and   Zerilli equations. We will see that second-order effects arise necessarily when expressing gauge-invariant
tensor modes constructed from the TT gauge through the first-order gravitational wave strains calculated in the RW  gauge. 
Of course, a precise estimation of  these effects goes beyond the scope of this note, but it is the natural step to take in the near future. 
 
The note is organized as follows. In Section II we discuss how to construct gauge-invariant second-order tensor perturbations. In Section III we deal with the second-order effects introduced by the measurement operation, while in Section IV we devote our attention to the second-order effects from the gauge-invariance. Section V contains our conclusions. The paper is supplemented by an Appendix which offers a similarity with the St\"uckelberg mechanism.

\section{Gauge-invariant second-order tensor perturbations}
\noindent
In this section we summarize how to construct second-order gauge-invariant tensor modes. The expert reader can skip this section.

Before launching ourselves in technicalities, let us start with some general remarks.  Firstly, let us point out that it does not exist a unique way to  construct    gauge-invariant tensor modes.  Gauge-invariant objects,  not depending  upon the   coordinate  definition in a  given gauge, can be defined. 
For instance, the  tensor modes at first-order  are gauge independent, since they remain  the same
in all gauges. On  the contrary,  the gravitational potential is  gauge-dependent since it  changes in different  time slicings.     A gauge-invariant combination can be constructed from the gravitational potential, but it is not unique. 
There is  an infinite number of ways of making a gravitational  potential  gauge-invariant and  what is the best gauge one should start from  to compute the actual observables  depends on the  measurement which is performed. Similarly, for the tensor modes at second-order there is not a unique way to render them gauge-invariant, the starting point being dependent upon the of observation one performs and, often, on  the comparison between the measured quantity and the theoretical prediction (analytical or numerical). 
 As we argued in the introduction, the TT and the RW gauges  play a special role, and in the following we will focus our attention on those gauges. Let us first though proceed in full generality. 
 
 \subsection{Metric transformation}
\noindent
Under a  generic   coordinate transformation of the form
\begin{equation}
\label{ct}
	x^\mu \to \wt x^\mu = x^\mu + \xi^\mu
	\qquad \text{with} \qquad 
	\xi^\mu \equiv \lp \alpha, \xi^i\rp,
\end{equation}
a generic  metric transforms as 
\begin{eqnarray}
g_{\mu\nu}\to \wt g_{\mu\nu}=\bar{g}_{\mu\nu}+\delta g_{\mu\nu},
\end{eqnarray}
where 
\begin{eqnarray}
\delta g_{\mu\nu}=\xi_{\mu;\nu}+\xi_{\nu;\mu},
\end{eqnarray}
and $\bar{g}_{\mu\nu}$ is the background unperturbed metric.  A semicolon denotes covariant differentiation.
The coordinate transformations in Eq. (\ref{ct}) are not considered to be infinitesimal, but it can also  be  finite. In that case, they can be expanded in terms of a fictitious, bookkeeping parameter $\epsilon$, which we will suppress in the following. 
Therefore,   Eq. (\ref{ct}) is written to second-order as \cite{Bruni:1996im} 
\begin{eqnarray}
\label{ct2}
	x^\mu \to \wt x^\mu = x^\mu + \xi_1^\mu+\frac{1}{2}\left(\xi_{1,\nu}^\mu \xi_1^\nu+\xi_2^\mu\right),
\end{eqnarray}
where  we have expanded 
\begin{eqnarray}
\xi^\mu=\xi_1^\mu +\frac{1}{2}\xi_2^\mu.  \label{xi}
\end{eqnarray}
Expanding the metric to second-order as 
\begin{eqnarray}
g_{\mu\nu}=\bar{g}_{\mu\nu}+\delta g_{\mu\nu}
\end{eqnarray}
where 
\begin{eqnarray}
\delta g_{\mu\nu}=
\delta_1g_{\mu\nu}+\frac{1}{2}\delta_2 g_{\mu\nu},
\end{eqnarray}
we find that the first and second-order metric perturbations $\delta_1g_{\mu\nu}$ and $\delta_2 g_{\mu\nu}$, transform as 
\begin{align}
\wt {\delta_1}g_{\mu\nu}&=\delta_1g_{\mu\nu}+
\bar{g}_{\mu\nu,\lambda}\xi_1^\lambda+\bar{g}_{\kappa\nu}\xi_{1,\mu}^\kappa+\bar{g}_{\mu\lambda}\xi_{1,\nu}^\lambda,  \label{ctm1}\\
\wt {\delta_2}g_{\mu\nu}&=\delta_2g_{\mu\nu}
+\bar{g}_{\mu\nu,\lambda}\xi^\lambda_2
+\bar{g}_{\mu\lambda}\xi^\lambda_{2~,\nu}
+\bar{g}_{\lambda\nu}\xi^\lambda_{2~,\mu}
+2\Big(
\delta_1 g_{\mu\nu,\lambda}\xi^\lambda_1
+\delta_1 g_{\mu\lambda}\xi^\lambda_{1~,\nu}
+\delta_1 g_{\lambda\nu}\xi^\lambda_{1~,\mu}
\label{1st}
 \\
&+
\bar{g}_{\mu\lambda,\alpha} \xi^\alpha_1\xi^\lambda_{1~,\nu}
+\bar{g}_{\lambda\nu,\alpha} \xi^\alpha_1\xi^\lambda_{1~,\mu}
+\bar{g}_{\lambda\alpha}  \xi^\lambda_{1~,\mu} \xi^\alpha_{1~,\nu}
\Big)+\bar{g}_{\mu\nu,\lambda\alpha}\xi^\lambda_1\xi^\alpha_1
+\bar{g}_{\mu\nu,\lambda}\xi^\lambda_{1~,\alpha}\xi^\alpha_1
\nonumber \\
&+\bar{g}_{\mu\lambda}\left(
\xi^\lambda_{1~,\nu\alpha}\xi^\alpha_1
+\xi^\lambda_{1~,\alpha}\xi^\alpha_{1,~\nu}
\right)
+\bar{g}_{\lambda\nu}\left(
\xi^\lambda_{1~,\mu\alpha}\xi^\alpha_1
+\xi^\lambda_{1~,\alpha}\xi^\alpha_{1,~\mu}
\right). \label{2nd}
\end{align}
Knowing the transformation properties of the metric perturbations allow us to construct gauge-invariant quantities. 
We will demonstrate this for the case of flat Minkowski spacetime with background metric 
\begin{equation}
	\d s^2 = -  \d t^2 + \gamma_{ij}\, \d x^i \d x^j,
\end{equation}
where the $x^i$ are generic curvilinear coordinates (which later on we will take to be polar coordinates).
We can parametrise the  perturbed metric $\delta g_{\mu \nu}$
as 
\begin{align}\label{metricpert}
\delta g_{00} = -2 \phi, \qquad \delta g_{0i} = B_i, \qquad \delta g_{ij} = 2  C_{ij}.
\end{align}
In terms of the SO(3) subgroup of the full Poincar\'e isometry group of the Minkowski background,  the perturbed metric is decomposed  into  scalar-vector-tensor (SVT) components by defining 
\begin{align}
B_i = B_{,i} - S_i, \qquad C_{ij} = -\psi \delta_{ij} + E_{,ij} + F_{(i,j)} + \frac{1}{2}h_{ij}.
\end{align}
The tensor and vector degrees of freedom are defined to be divergence-free (transverse)  and traceless and they satisfy the conditions
\begin{equation}
	S_{i,i} = 0,
	\qquad
		F_{i,i} = 0,
		\qquad
		\text{and}
		\qquad
	h^i_{i}=	h_{ij,j} = 0.
\end{equation}
We now express all  quantities at first and second-order perturbations around the background as
\begin{subequations}
\begin{align}
	\phi &= \phi_1 + \frac{1}{2} \phi_2 + \dots
	\\
	\psi &= \psi_1 + \frac{1}{2} \psi_2 + \dots
	\\
	B &= B_1 + \frac{1}{2} B_2 + \dots
	\\
	E  &= E_1 + \frac{1}{2} E_2 + \dots
	\\
	S_i  &= S_{1i} + \frac{1}{2} S_{2i} + \dots
	\\
	F_i  &= F_{1i} + \frac{1}{2} F_{2i} + \dots
	\\
	h_{ij} &= h_{1ij} +\frac{1}{2} h_{2ij} + \dots.
\end{align}	
\end{subequations}
We  express the vector $\xi^\mu$ as 
\begin{equation}
\label{a}
	\xi^\mu = \lp \alpha_1+ \frac{1}{2}\alpha_2 , \xi^i_1 + \frac{1}{2}\xi^i_2\rp 
	\qquad 
	\text{with}
	\qquad 
	\xi^i_a= {\beta_{a,}}^{i} + \gamma_{a}^{i},
\end{equation}
where $a=\{1,2\}$ and  $\gamma_a^i$ are  divergence-free vectorial parameters ($\gamma^i_{,i}=0$).
Using the transformation of the metric Eq. (\ref{1st}) for the first-order quantities, we find that the  first-order gauge transformations are given by \cite{Malik:2008im}
\begin{subequations}
\label{gt1o}
\begin{align}
\label{transphi1}
\widetilde {\A1} =& \A1 +\dot{\alpha_1},\\
\label{transpsi1}
\widetilde \psi_1 =& \psi_1,\\
\label{transB_1}
\widetilde B_1 =& B_1-\alpha_1+\dot{\beta_1},\\
\label{transE_1}
\widetilde E_1 =& E_1+\beta_1,\\
\label{transS1}
\widetilde {S_{1}^{~i}} =& S_{1}^{~i}-{\dot{\gamma_1}}^i, \\
\label{transF1}
\widetilde {F_{1}^{~i}} =& F_{1}^{~i}+\gami1,\\
\widetilde h_{1ij} =& h_{1ij}\,.
\end{align}
\end{subequations}
Similarly, using Eq. (\ref{2nd}), we find that at second-order the gauge transformation of the components of the metric perturbation can be written as  
\begin{subequations}
\begin{align}
\label{transphi2}
\wt {\A2} &= \A2+\dot{\alpha_2}
+\alpha_1\big(\ddot{\alpha_1} +2\dot{\phi_1}\big)
+2\dot{\alpha_1}\big(\dot{\alpha_1}+2\phi_1\big)
+\xi_{1k}
\big(\dot{\alpha_1}+2\phi_1\big)_{,}^{~k}
+\dot{\xi}_{1k}\big(\alpha_{1,}^{~k}-2B_{1k}-{\dot{\xi}_1^k}\big), \\
\label{transpsi2}
\wt\psi_2&=\psi_2-\frac{1}{4}\X^k_{~k}
+\frac{1}{4}\nabla^{-2} \X^{ij}_{~~,ij},
\\
\label{transB2}
\widetilde B_{2} &=B_{2}-\alpha_2+\dot{\beta_2} +\nabla^{-2} \XB^k_{~,k},
\\
\label{transE_2}
\wt E_2&=E_2+\beta_2+\frac{3}{4}\nabla^{-2}\nabla^{-2}\X^{ij}_{~~,ij}
-\frac{1}{4}\nabla^{-2}\X^k_{~k},
\\
\label{transS2}
\widetilde S_{2i}&=S_{2i}-\dot{\gamma}_{2i}-\XB_i+\nabla^{-2}\XB^k_{~,ki},
\\
\label{transFi2}
\wt F_{2i}&= F_{2i}+\gamma_{2i}
+\nabla^{-2}\X_{ik,}^{~~~k}-\nabla^{-2}\nabla^{-2}\X^{kl}_{~~,kli},
\\
\label{transhij2}
\wt h_{2ij}&= h_{2ij}+\X_{ij}
+\frac{1}{2}\left(\nabla^{-2}\X^{kl}_{~~,kl}-\X^k_{~k}
\right)\delta_{ij}
+\frac{1}{2}\nabla^{-2}\nabla^{-2}\X^{kl}_{~~,klij}\nonumber 
\\
&
+\frac{1}{2}\nabla^{-2}\X^k_{~k,ij}
-\nabla^{-2}\left(\X_{ik,~~~j}^{~~~k}+\X_{jk,~~~i}^{~~~k}
\right).
\end{align}
\end{subequations}
We have defined the  vector $\XB_i$ and the tensor $\X_{ij}$, which both depend only on the square of the  first-order quantities
\begin{align}
\label{defXBi}
\XB_i
&\equiv 
2\Big(
\dot{B}_{1i}\alpha_1
+B_{1i,k}\xi_1^k-2\phi_1\alpha_{1,i}+B_{1k}\xi_{1,~i}^k
+B_{1i}\dot{\alpha_1}+2 C_{1ik}{\dot{\xi}_{1}^k}
 \Big)-\alpha_{1,k}\xi_{1,i}^k\nonumber\\
&
+\dot{\alpha_1}\Big(\dot{\xi}_{1i}-3\alpha_{1,i}\Big)
+\alpha_1\Big(\ddot{\xi}_{1i}-\dot{\alpha}_{1,i}\Big)+{\dot{\xi}_{1}^k}\left(\xi_{1i,k}+2\xi_{1k,i}\right)
+\xi_{1}^k\Big(\dot{\xi}_{1i,k}-\alpha_{1,ik}\Big),
\end{align}
and 
\begin{align}
\label{Xijdef}
\X_{ij}&\equiv
4\Big(\alpha_1\dot{C}_{1ij}
+C_{1ij,k}\xi_{1}^{~k}+C_{1ik}\xi_{1~~,j}^{~k}
+C_{1kj}\xi_{1~~,i}^{~k}\Big)
+2\Big(B_{1i}\alpha_{1,j}+B_{1j}\alpha_{1,i}\Big)
\nonumber\\
&
-2\alpha_{1,i}\alpha_{1,j}+2\xi_{1k,i}\xi_{1~~,j}^{~k}
+\alpha_1\Big( \dot{\xi}_{1i,j}+\dot{\xi}_{1j,i} \Big)
+\Big(\xi_{1i,jk}+\xi_{1j,ik}\Big)\xi_{1}^{~k}+\xi_{1i,k}\xi_{1~~,j}^{~k}
\nonumber\\
&+\xi_{1j,k}\xi_{1~~,i}^{~k}
+\dot \xi_{1i}\alpha_{1,j}+\dot \xi_{1j}\alpha_{1,i}.
\end{align}
\subsection{Construction of second-order gauge-invariant tensor modes}
\noindent
Let us remind the reader how to construct gauge-invariant quantities from   a particular gauge \cite{Bartolo:2004if,Malik:2008im}.   Choosing a gauge is equivalent to pick up a  vector $\xi^\mu$ such that  certain conditions are satisfied by the metric. This enforces  the parameters  $\alpha$ and $\xi^{i}$ to be expressed in terms of the perturbation fields $(\delta g)$ or some combination thereof. Then,  the  particular form of  $\alpha (\delta g)$ and $\xi^{i}(\delta g)$ used to fix the gauge can be employed to perform a general gauge transformation of the original fields so that the new transformed fields are now  gauge-invariant quantities.  

Let us illustrate  how  this procedure  works for the first-order  scalar potentials $\phi_1$ (see also Appendix A). We can choose to set the parameters $\alpha_1$ and $\beta_1$ to go to a gauge where 
$\widetilde B_1=\widetilde E_1=0$. From Eqs. (\ref{transB_1}) and (\ref{transE_1}), this will determine $\beta_1=-E_1$ and $\alpha_1=B_1-\dot E_1$. Inserting these choices into Eq. (\ref{transphi1}) we find a  gauge-invariant expression for the gravitational potential
\begin{align}\label{sgi1}
 \phi_{1}^\GI & \equiv \phi_{1} +\dot B_1-\ddot E_1.
 \end{align}
A similar procedure can be used to define gauge-invariant
second-order transverse-free and traceless perturbation which, as we will discuss later on, is to be  identified with the tensor modes in the TT gauge. Using the gauge transformation properties of the tensor as in  Eq.~\eqref{transhij2} one finds \cite{Matarrese:1997ay,Malik:2008im}\footnote{Non-local terms in the definition of the gauge-invariant second-order tensor modes are there to ensure that the modes are transverse and traceless. They disappear in the "projected" equation of motion.}
\begin{align} 
h_{2ij}^\GI& \equiv h_{2ij}+\X_{ij}^{{\text{\tiny GC}}}
+\frac{1}{2}\left(\nabla^{-2}\X^{{\text{\tiny GC}} kl}_{~~~,kl}-\X^{{\text{\tiny GC}} k}_{~~~k}
\right)\delta_{ij}
+\frac{1}{2}\nabla^{-2}\nabla^{-2}\X^{{\text{\tiny GC}} kl}_{~~~~,klij}
\nonumber \\
&
+\frac{1}{2}\nabla^{-2}\X^{{\text{\tiny GC}} k}_{~~~k,ij}
-\nabla^{-2}\left(\X_{ik,~j}^{{\text{\tiny GC}} k}+\X_{jk,~i}^{{\text{\tiny GC}} k}
\right), \label{h2gi}
\end{align}
where now
\begin{align}
\X_{ij}^{\text{\tiny GC}} &\equiv 
4\Big(\alpha_1^{\text{\tiny GC}} \dot C_{1ij}
+C_{1ij,k}\xi_{1}^{{\text{\tiny GC}} k}+C_{1ik}\xi_{1~~,j}^{{\text{\tiny GC}} k}
+C_{1kj}\xi_{1~~,i}^{{\text{\tiny GC}} k}\Big)
+2\left(B_{1i}\alpha_{1,j}^{\text{\tiny GC}} +B_{1j}\alpha_{1,i}^{\text{\tiny GC}} \right)
\nonumber\\
&
-2\alpha_{1,i}^{\text{\tiny GC}} \alpha_{1,j}^{\text{\tiny GC}} +2\xi_{1k,i}^{\text{\tiny GC}} \xi_{1~~,j}^{{\text{\tiny GC}}  k}
+\alpha_1^{\text{\tiny GC}} \left( \dot \xi_{1i,j}^{{\text{\tiny GC}} }+\dot\xi_{1j,i}^{{\text{\tiny GC}} } \right)
+\left(\xi_{1i,jk}^{\text{\tiny GC}} +\xi_{1j,ik}^{\text{\tiny GC}} \right)\xi_{1}^{{\text{\tiny GC}} k}
\nonumber\\
&+\xi_{1i,k}^{\text{\tiny GC}} \xi_{1~~,j}^{{\text{\tiny GC}}  k}+\xi_{1j,k}^{\text{\tiny GC}} \xi_{1~~,i}^{{\text{\tiny GC}}  k}
+\dot \xi_{1i}^{{\text{\tiny GC}} }\alpha_{1,j}^{\text{\tiny GC}} +\dot \xi_{1j}^{{\text{\tiny GC}}} \alpha_{1,i}^{\text{\tiny GC}}
\label{h2gi1}
\end{align}
in terms of the fields $\alpha_{1}^\text{\tiny GC} (\delta g)$ and $\xi_{1i}^\text{\tiny GC}(\delta g)$. The label ($\text{GC}$)   reminds  that the corresponding quantity is specified by solving a given (and arbitrary)  gauge condition.  
Different gauge conditions  give rise to  different gauge-invariant  quantities by using this procedure. 
However, there are also gauge-independent quantities, that is quantities that are independent of the gauge used. For example, the first-order transverse-traceless part $h_{1ij}$ of the tensor perturbation $C_{ij}$ (like $\psi_1$) is gauge-independent: it is invariant in any gauge. On the other hand, 
$h_{2ij}^\GI$ is gauge-invariant,  but it depends on the gauge used since it depends on the parameter in  $\xi^{\text{\tiny GC}\mu} (\delta g)$. 

What is more relevant for us is that the gauge-invariant second-order tensor mode automatically contains in its definition terms of the form (first-order)$^2$ which come from two sources: those contain in the intrinsically second-order quantity $ h_{2ij}$ which are determined by the dynamics (i.e. the merger of two BHs) and those which are explicitly present in the tensor $\X_{ij}^{{\text{\tiny GC}}}$ whose introduction is necessary and unavoidable to define the gauge-invariant second-order tensor mode.

\section{The nonlinearities from the measurement of GWs}\label{secm}
\noindent
In this section we discuss  the presence of the nonlinearities coming from the measurement operation of the GWs.
In interferometer experiments of arms of length $L$ the measurement is done by sending  photons to the mirrors located  and observing the modulation in power recorded because of  the different time shifts $\Delta t_{\text{\tiny A},\text{\tiny B}}$ acquired in the different travel paths along the arms A and B. 
Since for space-based observatories (like LISA) the frequency $\omega$ of the GWs  is such that  $\omega L={\cal O}(1)$, a single reference frame may not be adopted for which the whole apparatus is described by an (approximately) flat metric in the presence of the GW. On the contrary, a completely general relativistic framework has to be used. In this respect, 
 the most suitable coordinate system turns out to be the TT frame 
where  the coordinates in the positions of the mirrors (for a thorough discussion about the virtues of using the TT frame see Ref. \cite{book}). 

Here we briefly summarises how to define the TT gauge at first-order. The conditions to impose are $\delta g_{00}=\delta g_{0i}=0$, which at first-order lead to 
\begin{align}
\alpha_{1}^\syn
&=-\llp \int \phi_1 \d t-\C_1(\bm{x})\rrp,\label{sync1}\\
\beta_{1}^{\syn}&=\int\left( \alpha_{1}^{\syn}-B_1\right) \d t
+\hat\C_1(\bm{x}),\label{sync2} \\
 \gamma_{1i}^{\syn} &= \int S_{1i} \d t+ \hat\C_{1i}(\bm{x}).
 \label{syn3}
 \end{align}
The time-slicing is fully determined   once two   arbitrary functions of the spatial coordinates $\C_1(\bm{x})$ and $\hat\C_{1}(\bm{x})$ are fixed. Furthermore, there is an arbitrary  3-vector $\hat\C_{1i}$ (with 2 independent components) depending upon
the choice of spatial coordinates on an initial hypersurface. 
These extra four degrees of freedom  can be further fixed by imposing the  transverse-traceless (TT) condition 
\begin{eqnarray}
{{C_1}^{ik}}_{,k}=C^k_{1k}=0. \label{TT}
 \end{eqnarray} 
 A similar procedure can be performed at second-order. 
  In the TT gauge the second-order metric (by writing only the GW perturbation and using cartesian coordinates) reads 
\begin{equation}
	\d s^2 =
	- \d t^2 
	+ \lp \delta_{ij} + h^\text{\tiny TT}_{1ij} + \frac{1}{2}h^\text{\tiny TT}_{2ij} \rp\d x^i \d x^j,
\end{equation}
where $ h^\text{\tiny TT}_{ij}$ is the transverse and traceless components of the $C_{ij}$ in Eq. (\ref{metricpert}).
 
The  effect of a  GW passing through the interferometer is captured by measuring the  proper times at the interferometers. Photons travel along the arms following the geodesic equation 
$\d s^2 = 0$ and we find up to second-order that the time shift reads (along  two arms)
\begin{align}
	\label{tdTT}
	\Delta t_{\text{\tiny A},\text{\tiny B}} & = L_{\text{\tiny A},\text{\tiny B}} - \int_{t_0}^{t_0+2 L_{\text{\tiny A},\text{\tiny B}}}
	\d t
	 \lp - \frac{1}{2}h^\text{\tiny TT}_{1ij}  +\frac{3}{8}(h^\text{\tiny TT}_{1ij})^2 - \frac{1}{4} h^\text{\tiny TT}_{2ij} \rp _{i=j=1,2}.
\end{align}
From these simple arguments we see that second-order effects of the QNMs   enter not only through the intrinsically second-order quantity $h^\text{\tiny TT}_{2ij}$, but also through the explicit (first-order)$^2$ term $3(h^\text{\tiny TT}_{1ij})^2/8$ in the time shifts. We expect that rendering the time shifts fully gauge independent will cause the appearance of other 
(first-order)$^2$ terms which must be taken into account in the gravitational waveform analysis.

\section{The nonlinearities from gauge invariance}
\noindent
In  Section II  we showed that at first-order the tensor  part of the metric perturbation $h_{1ij}$ is invariant under coordinate transformation and that its second-order counterpart is not. This means that at second-order any  gauge-invariant tensor quantity  will contain pieces which are (first-order)$^2$ which will eventually contribute through the QNMs to the time shifts measured in the interferometers.  

We also learned in Section II that the TT gauge is preferable when dealing with  observations. It is therefore natural to construct a gauge-invariant quantity starting from the TT gauge as we described in Section II. 
The procedure will be therefore the following. Starting from a fully generic gauge, one chooses the parameters of the coordinate transformations (\ref{a}) by setting $\delta g_{00}=\delta g_{0i}=0$ as well as the transverse and traceless condition on the spatial part of the metric. With this procedure we arrive at the generic gauge-invariant expression for the tensor mode at second-order
(\ref{h2gi}). The transverse and traceless condition, by construction, does not depend  neither on $h^\text{\tiny TT}_{1ij}$ nor on  $h^\text{\tiny TT}_{2ij}$. Therefore, 
the quadratic pieces in the gauge-invariant second-order tensor mode does not  contain explicit terms of the form ${\cal O}[(h^\text{\tiny TT}_{1ij})^2]$. Terms quadratic in $h^\text{\tiny TT}_{1ij}$ will appear only in $h^\text{\tiny TT}_{2ij}$ once its dynamics is solved at second-order. In other words
\be
\label{simp}
h_{2ij}^\GI\big|_{\text{\tiny in the TT gauge}} =h^{\text{\tiny TT }}_{2ij}.
\ee
In order to compute $h_{2ij}^\GI$ one has therefore various  choices.
Either one computes it in the TT gauge directly and writes  an equation
to calculate the evolution of $h^{\text{\tiny TT }}_{2ij}$,  or  otherwise, one can 
calculate  $h_{2ij}^\GI$ in another gauge. Alternatively, one can adopt a sort of mixed procedure where one calculates the second-order evolution in a given gauge and then go to the TT gauge. 
We will pursue this third option here.  The  so-called RW gauge  is convenient to solve for the  
Schwarzschild perturbations through the RW and   Zerilli equations, which describe axial and polar perturbations, respectively. 
We first write the TT part of the perturbation in polar coordinates as
\begin{eqnarray}
{h}_{1\mu\nu}^\text{\tiny TT}(t,r,\theta,\phi)= \sum_{\ell=2}^\infty \sum_{m=-\ell}^{\ell}
{h}_{\ell m}^\text{\tiny TT}(t,r){\big(t^{(E2)}_{\ell m}\big)}_{\mu\nu}, 
\label{hax}
\end{eqnarray}
where 
\begin{eqnarray}
\Big(t^{(E2)}_{\ell m}\Big)_{\mu\nu}&=&\frac{r^2}{2}\left(
\begin{array}{cccc}
   0  &0& 0& 0
    \\
   0&0&0&0\\
   0&0&W&X\\
   0&&X&-\sin^2\theta\,W
\end{array}\right)Y_{\ell m},\nonumber\\
W&=&\partial_\theta^2-\cot\theta\partial_\theta-\frac{1}{\sin^2\theta}\partial^2_\phi,\,\,\,\,X=2(\partial_\theta\partial_\phi-\cot\theta\partial_\phi).
\end{eqnarray}
 $Y_{\ell,m}$ are the spherical harmonics.
Following Ref. \cite{Nakano:2007cj}, we can write the fundamental modes  at infinity in the TT gauge at first- and second-order as (we set $G_N=1$)

\begin{eqnarray}
h^\text{\tiny TT}_{1,(2,\pm 2)}&\simeq &\frac{1}{r}\psi_{1,(2,\pm 2)}(t,r)\nonumber\\
h^\text{\tiny TT}_{2,(4,\pm 4)}&\simeq &\frac{1}{r}\psi_{2,(4,\pm 4)}(t,r)-i\frac{\sqrt{70}\cdot 51}{1512\cdot 2\sqrt{\pi}}\frac{(\omega_{2,\pm 2})}{r}\left(\psi_{1,(2,\pm 2)}(t,r)\right)^2,
\end{eqnarray}
where the $\psi_{1,2(\ell,m)}$ satisfy the corresponding first- and second-order Zerilli equation in the RW gauge and the second-order part has been properly regularized. At first-order, for instance, the equation reads

\begin{eqnarray}
&&\left(-\frac{\partial^2}{\partial t^2}+\frac{\partial^2}{\partial r_*^2}-V_Z(r)\right)\psi_{1,(\ell,m)}(t,r)=0,\nonumber\\
V_Z(r)&=& \left(1-\frac{2M}{r}\right)\frac{2\lambda^2(\lambda+1)r^3+6\lambda^2M r^2+18\lambda M^2 r+18M^3}{r^3(\lambda r+3M)^2},\,\,\,\,\lambda=\frac{1}{2}(\ell-1)(\ell+2),\nonumber\\
&&\hskip 2cm r^*=r+2M\ln\left(\frac{r}{2M}-1\right),
\end{eqnarray}
and the reader can find in Ref. \cite{Nakano:2007cj} the equation for $\psi_{2,(\ell,m)}$ with its corresponding (regularized) second-order source. 
Taking the approximation that the QNMs are produced around the peak of the (minus of the) Zerilli potential $V_Z$, at $(r^{\ell=2}_\text{\tiny pk}/M)\simeq 3.1$ and  $(r^{\ell=4}_\text{\tiny pk}/M)\simeq 3.05$ and  using the fact that for the $n=0$ mode  the fundamental wavefunctions are
\begin{eqnarray}
\psi_{1,(2,\pm 2)}(t,r)&\simeq& A_{1,(2,\pm 2)}e^{-i\omega_{(2,\pm 2)}t}e^{-z^2/4},\nonumber\\
\psi_{1,(4,\pm 4)}(t,r)&\simeq& A_{1,(4,\pm 4)}e^{-i\omega_{(4,\pm 4)}t}e^{-z^2/4},\nonumber\\
z&\simeq& (4 k)^{1/4}e^{3\pi i/4}(r^*-r^*_\text{\tiny pk}),\,\,\,\,r^*_\text{\tiny pk}(\ell=2)\simeq 1.9,\,\,\,\,r^*_\text{\tiny pk}(\ell=4)\simeq 1.76, \nonumber\\
k&=&-V''_Z(r^*_\text{\tiny pk})/2, 
\end{eqnarray}
a saddle-point approximation specifies  the corresponding amplitude to be
\be
A_{2,(4,\pm 4)}=\frac{A^2_{1,(2,\pm 2)}}{M}\left[0.05-0.08(\omega_{2,\pm 2}M)+0.24(\omega_{2,\pm 2}M)^2-2.9(\omega_{2,\pm 2}M)^4\right].
\ee
Taking $(\omega_{2,\pm 2}M)\simeq 0.37$, one finally finds that $A_{2,(4,\pm 4)}\simeq 0.06A^2_{1,(2,\pm 2)}/M$. Since one expects  $A_{1,(4,\pm 4)}={\cal O}(0.1)\,A_{1,(2,\pm 2)}/M$ \cite{nl2}, we indeed see that the second-order  mode for $\ell=4$ can indeed be of the same order of magnitude of the corresponding linear mode for sizeable amplitudes of the fundamental mode $\ell=2$. It will be interesting to see if some hidden symmetries may explain the nonlinearties of the Schwarzschild BH along the lines of Ref. \cite{Kehagias:2023ctr} for extremal Kerr BHs.

%
\noindent

\begin{acknowledgments}
\noindent
We thank V. De Luca and G. Franciolini for very useful discussions.  A.R. acknowledges financial support provided by the Boninchi Foundation. 

\end{acknowledgments}

\appendix

\section{The  St\"uckelberg procedure}
\noindent
The construction in Section II can more easily be illustrated for the case of Maxwell theory. Here the gauge transformations are the $U(1)$ transformations of the vector potential $A_\mu$, which read 
\begin{eqnarray}
A_\mu\to \wt A_\mu=A_\mu+\phi_{,\mu}.  \label{gm}
\end{eqnarray}
We may express $A_\mu$ as 
\begin{eqnarray}
A_\mu=(A_0,C_i+F_{,i})\qquad 
	\text{with}
	\qquad C^i_{,i}=0,
\end{eqnarray}
so that the gauge transformation is given by 
\begin{align}
\wt A_0&=A_0+\dot{\phi}, \nonumber \\
\wt F&=F+\phi, \nonumber \\
\wt C_i&= C_i.  \label{trg}
\end{align}
Therefore, like the case of metric perturbations, the transverse part of the gauge potential does not change under $U(1)$ gauge transformations. Similarly, the scalars $A_0$ and $F$ change according to  (\ref{trg}).  Let us now choose a particular gauge, for example  the temporal gauge
\begin{eqnarray}
\wt A_0=0. 
\end{eqnarray}
Under a gauge transformation we get that 
\begin{eqnarray}
\wt A_0=A_0+\dot{\phi}=0, 
\end{eqnarray}
from where we find that 
\begin{eqnarray}
\dot{\phi}=-A_0. 
\end{eqnarray}
The solution of the above equation provides what we call 
$\phi^\text{\tiny GC}$, or in other words
\begin{eqnarray}
\phi^\text{\tiny GC}=-\int^tA_0(\tau,\vec{x})\d \tau.
\end{eqnarray}
Having specified the gauge parameter $\phi^\text{\tiny GC}$, it is easy to construct gauge-invariant quantities. For example, following the discussion above, the gauge-invariant scalar is now
\begin{eqnarray}
F^{\GI}=F_{\rm tm}^{\GI}=F+\phi^\text{\tiny GC}=F-\int^tA_0(\tau,\vec{x})\d \tau. 
\label{FI}
\end{eqnarray}
The scalar $F^{\GI}$ is indeed gauge-invariant since under a gauge transformation of the form (\ref{gm}) with a gauge parameter $\chi$, we get that 
\begin{eqnarray}
\wt A_0=A_0+\dot{\chi}, \qquad \wt F=F+\chi,
\end{eqnarray}
and therefore, from Eq. (\ref{FI}), the scalar $F^{\GI}$ is gauge-invariant. In addition, the transverse part $C_i$ of $A_i$ does not change at all. If we use another gauge, let say the axial $A_3=0$, we can repeat the above construction. In this case though,the gauge-invariant scalar will be 
\begin{eqnarray}
F_{\rm ax}^{\GI}=F-\int^{x_3} A_3(t,x_a,y)\d y, \qquad a=(1,2)
\end{eqnarray}
but still $\wt C_i=C_i$. In other words, changing the gauge, the gauge-invariant quantities change accordingly, while gauge-independent quantities do not change from gauge to gauge. Of course, one recognises in the above construction the  St\"uckelberg trick to build gauge-invariant quantities.

\bibliographystyle{JHEP}
\bibliography{draft}

\end{document}